\documentclass[fleqn]{mn2e}
\usepackage{graphicx}
\usepackage{mynatbib}
\usepackage{amsmath}

\newcommand{\msun}{{\rm M_{\odot}}}

\newcommand{\lambdabar}{                          
  \hbox{\raise 0.1em
  \hbox to 0pt {--\hss}$\lambda$}}

\renewcommand{\epsilon}{\varepsilon}              

\newcommand{\TTMT}{thermal--timescale mass transfer}

\newcommand{\newblock}{}

\title{AE Aquarii: how CVs descend from supersoft binaries}

\author[K.~Schenker et al.]
{K.~Schenker$^{1}$, A.R.~King$^{1}$, U.~Kolb$^{2}$, G.A.~Wynn$^{1}$, Z.~Zhang$^{1,2,3}$\\
           $^1$ Theoretical Astrophysics Group, Department of Physics and Astronomy,
	        University of Leicester, Leicester, LE1 7RH, U.K.\\
           $^2$ Department of Physics and Astronomy, The
		Open University, Walton Hall, Milton Keynes, MK7 6AA, U.K.\\
           $^3$ National Astronomical Observatories, Chinese Academy
		of Sciences, Beijing 100012, P.R. China}
\date{Accepted. Received}
\pagerange{\pageref{firstpage}--\pageref{lastpage}}
\pubyear{2002}

\begin{document}
\label{firstpage}

\maketitle


\begin{abstract}
AE~Aquarii is a propeller system. It has the shortest spin period
among cataclysmic variables, and this is increasing on a $10^7$~yr
timescale. Its UV spectrum shows very strong carbon depletion vs
nitrogen and its secondary mass indicates a star far from the
zero--age main sequence. We show that these properties strongly
suggest that AE~Aqr has descended from a supersoft X--ray binary. We
calculate the evolution of systems descending through this channel,
and show that many of them end as AM~CVn systems. The short spindown
timescale of AE~Aqr requires a high birthrate for such systems,
implying that a substantial fraction of cataclysmic variables must
have formed in this way. A simple estimate suggests that this fraction
could be of order one--third of current CVs. We emphasize the
importance of measurements of the C/N abundance ratio in CVs,
particularly via the {\sc Civ} 1550/{\sc Nv} 1238 ratio, in determining how
large the observed fraction is.

\end{abstract}



\section{Introduction}

AE~Aqr is one of the most distinctive cataclysmic variables (CVs). It
has a long orbital period of 9.88~hr \citep{Welsh_etal:637}, and the
shortest coherent pulse period (33~s), increasing on a timescale
$\sim 10^7$~yr \citep{Jager_etal:1018}. Doppler tomography reveals the
apparent absence of an accretion disc. This has led to its
interpretation as a `propeller' system \citep{Wynn_etal:624}:
the white dwarf is apparently expelling
centrifugally the matter transferred from the companion. 
To arrive at such a highly non--equilibrium spin rate, the mass
transfer rate in the recent ($10^7$~yr) past must have
been much higher than its current value 
$\dot{M}_2 \sim$ few $\times \, 10^{-9} \, \msun {\rm yr}^{-1}$, 
and decreased on a still shorter timescale. 
IUE spectra of AE~Aqr \citep{Jameson_etal:1019} reveal the most extreme
{\sc Civ} to {\sc Nv} ratio of all CVs \citep{Mauche_etal:661}, probably
indicating strong carbon depletion and thus CNO cycling. The short
spin period makes the system effectively a double--lined spectroscopic
binary, with masses $M_1 \simeq 0.89 \pm 0.23 \, \msun$ for the white
dwarf (WD) and $M_2 \simeq 0.57 \pm 0.15 \, \msun$ for the donor star
\citep{Welsh_etal:625,Casares_etal:623}. The donor star is of spectral
type K5V \citep{Welsh:668} which is too late for the given orbital
period and too early for the measured $M_2$ if it is on the main
sequence.

We show here that all of these properties are consistent with the idea
that AE~Aqr descends from a supersoft X--ray binary, cf.\
\citet{Schenker+King:892}. The thermal--timescale mass transfer
characterizing such sources ends once the secondary/primary mass ratio
$q = M_2/M_1$ decreases sufficiently, leading to a rapid transition to
normal CV evolution driven by angular momentum losses, cf.\
\citet{King:19}. If the white dwarf is magnetic, as in AE~Aqr, the
high mass transfer rate in the thermal--timescale phase will have spun
it up to a short spin period. This must lengthen rapidly in order to
reach equilibrium with the lower transfer rate in the CV state, and
this spindown is what drives its propeller action. The very short
lifetime of the propeller phase implies a high birthrate for such
systems, comparable to those of known CVs. This in turn suggests that
systems descending from evolutions like AE~Aqr must constitute a large
fraction of all CVs.

The simple idea that all CVs form with essentially unevolved,
low--mass main sequence (MS) donors has been repeatedly challenged for
more than a decade \citep{Pylyser+Savonije:587,Pylyser+Savonije:588,
Baraffe+Kolb:613,Schenker:724,King+Schenker:891,Schenker+King:892}. 
If mass transfer starts with
mass ratio $q \ga 1$ the secondary's Roche lobe $R_{\rm L}$ tends to
shrink with respect to its thermal--equilibrium radius $R_{\rm
TE}$. Mass is therefore transferred on a thermal timescale ($\sim$ few
$\times 10^7$~yr), at rates high enough to sustain steady nuclear
burning on the material accreted by the white dwarf, and thus explain
(short-period) supersoft X-ray sources
\citep{Heuvel_etal:583,King_etal:640}.

After this relatively brief phase, $q$ becomes small enough that
$R_{\rm L}$ shrinks more slowly than $R_{\rm TE}$, and
thermal--timescale mass transfer ends. The systems either evolve off
to longer orbital periods, driven by nuclear evolution, or switch to
stable, angular-momentum--loss-driven mass transfer like ordinary
CVs. We will show that AE~Aqr is at the end of this transition phase
to a CV, accounting for its strange properties. AE~Aqr is passing
through this transition very rapidly; its whole evolution up to its
current state is short compared to the stable CV phase afterwards. In
this sense AE~Aqr is not unique at all: a significant fraction of CVs
must descend from similar evolutions.

In the next section we derive some rough estimates for the
potential progenitor system of AE~Aqr, using a simplified analytic
description of the orbital evolution during rapid mass transfer.
Sect.~3 studies the behaviour of single star
models under fixed mass loss of the order expected during such an
evolution. This allows us to identify the influence of varying stellar
parameters, and establish whether the observed properties of AE~Aqr
are compatible with this scenario. We present a set of full binary
evolution calculations in Sect.~4 and discuss them in Section~5.
Section~6 is the Conclusion.

\section{Simplified orbital evolution}

\begin{figure}
 \begin{center}
  \centerline{\includegraphics[clip,width=0.95\linewidth]{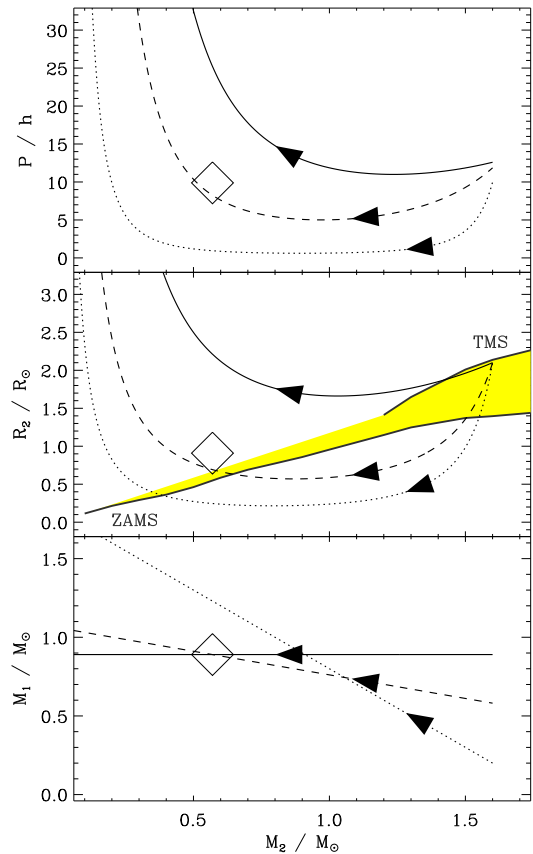}}
  \caption{Evolution of a model AE~Aqr progenitor system.  For the
  assumed parameters and $M_{\rm 1,now} = 0.89 \, \msun$, various
  tracks are shown, starting from a $1.6 \, \msun$ star which has
  almost reached its maximal MS radius.  The three panels show the
  evolution of the orbital period, secondary radius and WD mass for
  different cases (dotted line: $\eta = 1$ --- full line: $\eta = 0$
  --- dashed line: $\eta = 0.3$). The diamond marks the current
  position of AE~Aqr. Note that the conservative model ($\eta = 1$)
  cannot meet the requirements for $M_1$: even starting from the
  lowest plausible WD mass, it has grown well beyond the Chandrasekhar
  limit before reaching the current $P_{\rm orb}$ of AE~Aqr.
  The grey shaded area in the middle panel marks the radii of single
  stars during their main sequence life as labelled.
} 
  \label{fig:r_evol}
 \end{center}
\end{figure}

To get a first idea about the required initial parameters for
the progenitor system of AE~Aqr, we recall that during \TTMT\
additional angular momentum losses (magnetic braking \& gravitational 
radiation) are negligible compared to the changes of the Roche radius
resulting from mass transfer itself, e.g.\ \citet{King_etal:640}.
We define $\eta$ by
\begin{equation}
 \dot M_{1} =  -\eta \dot M_2
 \label{eta}
\end{equation}
as the fraction of the mass lost from the secondary that is actually
accreted on to the WD, while assuming that the rest leaves the system
with average specific angular momentum equal to the primary's, 
i.e.\ $\beta = 1$ in the nomenclature of \citet{King_etal:640}. From
their eq.~(10) we obtain the evolutionary tracks as 
\begin{equation}
 \frac{P}{P_i} = \left(\frac{M_{2i}}{M_2}\right)^3
		\left(\frac{M_i}{M}\right)^2
		\left(\frac{M_{1i}}{M_1}\right)^{3/\eta}
 \label{p}
\end{equation}
and 
\begin{equation}
 \frac{R_2}{R_{2i}} = \left(\frac{M_{2i}}{M_2}\right)^{5/3}
		\left(\frac{M_i}{M}\right)^{4/3}
		\left(\frac{M_{1i}}{M_1}\right)^{2/\eta}
 \label{r}
\end{equation}
respectively for $\eta > 0$. For the limit $\eta = 0$ these lead to the
exponential expressions given e.g.\ in \citet{King+Ritter:610}.

Equipped with these expressions we can estimate where analytically
computed curves for the evolution of radius and orbital period
intersect those describing mass loss under the assumption of permanent
thermal equilibrium (i.e.\ essentially along the MS). These
intersections mark the transition between the fast \TTMT\ (supersoft) and
CV--like angular--momentum--loss--driven evolution at lower mass
transfer rates.  Figure \ref{fig:r_evol} shows a set of curves for
different $\eta$
together with the shaded location of the MS framed by a pair of thick
curves marking the zero--age MS (ZAMS) and the turn--off MS (TMS).
  For the latter we sue the point of maximal radius prior to core hydrogen
  exhaustion. Note that this definition is not applicable to MS stars
  without convective cores (i.e.\ with masses below $\sim 1.2 \, \msun$).
The lowest panel shows the growth of the WD mass as the donor evolves
(from right to left) for various $\eta$.  The initial WD mass $M_{1i}$
is chosen so that the current position of AE~Aqr is reached with
masses of $(M_1,M_2) = (0.89 \, \msun,0.57 \, \msun)$ in accordance
with observation. A good fit should also pass through the correct
orbital period and radius (estimated by the corresponding period --
mean density relation). The assumption of fully conservative mass
transfer ($\eta = 1$) leads to much higher WD masses than observed,
and we therefore do not consider it further.

Using the curves shown
%
we derive first guesses for the system parameters of the progenitor
system, i.e.\ $M_{1i}$, $M_{2i}$, and $\eta$. We cannot
expect these estimates to be very precise, as the actual reaction of the system to
mass loss in this transition phase follows neither the Roche curves
nor the equilibrium curves (the secondary is out of thermal equilibrium).

Nevertheless we can estimate $M_{2i} \sim 2 \, \msun$ and $1.5 \la q_i
\la 3$, where the upper constraint derives from the occurence of
delayed dynamical instabilities (DDI, see \citet{King_etal:640} and
discussion section).  As explained above, conservative mass transfer
gives too high a WD mass. However a value $\eta \simeq 0.3$ is
plausible and gives an $M_1$ mildly increased over the
average field WD mass, in accordance with observations of
AE~Aqr.

\section{Single star calculations}

\begin{figure}
 \begin{center}
  \centerline{\includegraphics[clip,width=0.95\linewidth]{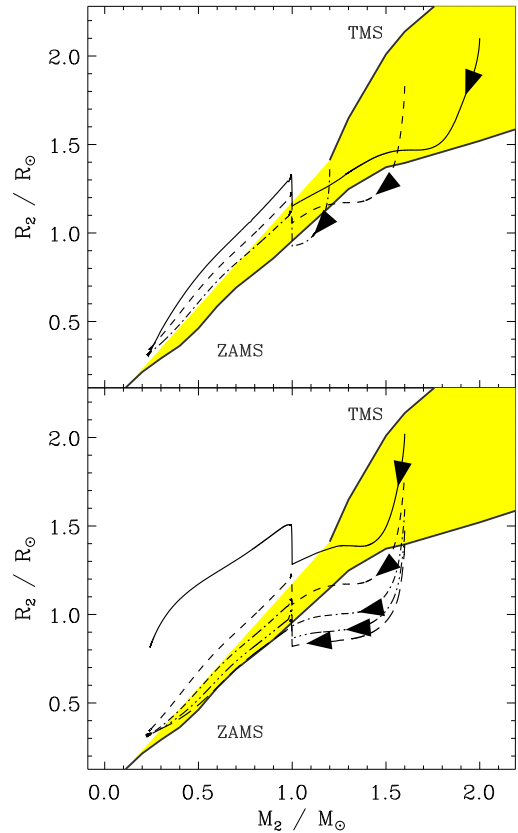}}
  \caption{Single star calculations with a mass loss rate according to 
         eq.(\protect\ref{switch}).
	 The upper panel shows the radius evolution of mass--losing
         stars with initially $1.2, 1.6, 2.0 \, \msun$ and an initial
	 age 0.75 of their MS lifetime. 
	 The lower panel illustrates the variation with age for a
	 $1.6 \, \msun$ star, covering an initial age of 
	 0, 0.25, 0.5, 0.75 and 1 MS-timescale from bottom to top.
	 Note that the strong deviation of the most evolved of the
	 five tracks results from ongoing nuclear evolution, which is
	 fast enough because the system was already close to entering
	 the Hertzsprung gap. The grey shaded area and additional
	 lines mark the location of the MS (as labelled). Thermal
	 timescale mass transfer was switched off at $M_2 = 1.0 \, \msun$ 
	 in each of the computations.
	} 
  \label{fig:zz}
 \end{center}
\end{figure}
\begin{figure}
 \begin{center}
  \centerline{\includegraphics[clip,width=0.95\linewidth]{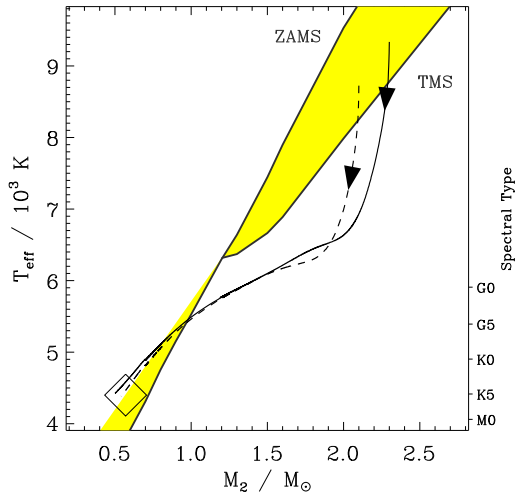}}
  \caption{Evolution of effective temperature (spectral type) with
        decreasing mass in the two best--fitting single star models S3
	with $M_{2i} = 2.1\, \msun$ (dashed line) and S4 with
	$2.3 \, \msun$ (full line). Details of the conversion
	from $T_{\rm eff}$ to SpT are discussed in the text.
	Thermal timescale mass transfer was switched off at
	$M_2 = 0.7 \, \msun$ in accord with the actual WD mass
	in AE~Aqr. Tracks end close to the current position of AE~Aqr,
	marked by a diamond. Note that the direction of evolution of
	single stars on the MS (shaded area) is towards lower 
	$T_{\rm eff}$ in more massive stars, but towards higher
	temperatures at lower masses.
	}
  \label{fig:zz_tem}
 \end{center}
\end{figure}
\begin{figure}
 \begin{center}
  \centerline{\includegraphics[clip,width=0.95\linewidth]{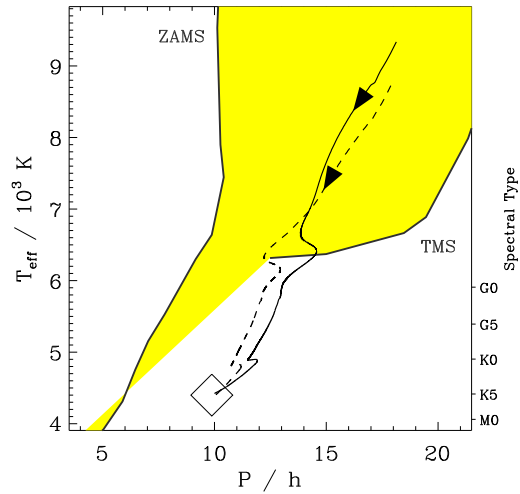}}
  \caption{Evolution of effective temperature (spectral type) with
    orbital period, for the same two tracks S3 \& S4 as in the
    previous figure.
    The large wiggles in orbital period during the
    early stage (at a SpT slightly above G0) are caused by the
    inconsistency introduced by an imposed mass transfer rate
    eq.(\ref{switch}), and the smaller ones near SpT K0 are
    directly connected with the jump in $\dot{M}_2$.
    The exact MS location (shaded area) in orbital period is of course
    a function of the primary mass, for which we have assumed the 
    canonical value of $M_1 = 0.89 \, \msun$ in this figure.}
  \label{fig:zz_tep}
 \end{center}
\end{figure}

We have shown that in principle a recent \TTMT\ supersoft phase could
explain AE~Aqr.  We now use full stellar models.
%
%
This determines the star's
reaction to mass loss, and thus whether the observed
spectral type (SpT) may be reached in this way. 
We simplify the mass transfer history as 
\begin{equation}
- \dot{M}_2 = 
 \begin{cases}
   M_2 / \tau_{\rm KH}                     & \forall M_2 >   M_{\rm sw} \\
   2 \times 10^{-9} \, \msun {\rm yr}^{-1} & \forall M_2 \le M_{\rm sw}
 \end{cases}
 \label{switch}
 \; 
\end{equation}
where $\tau_{\rm KH}$ is the secondary star's Kelvin--Helmholtz time,
thus mimicking \TTMT\ and the subsequent transition to an evolution
driven by angular momentum loss, as in ordinary CVs.  The switching
mass $M_{\rm sw}$ is chosen so that \TTMT\ ends when $q \simeq 1$,
i.e.\ it can be roughly identified with $M_1$.  A more thorough
discussion of the exact stability conditions in CVs is presented in
\citet{Webbink:702,Sobermann_etal:584}.  

The code we use in both the single star case of the last Section and
the binary calculations reported here is based on
\citet{Mazzitelli:197}, as adapted by \citet{Kolb+Ritter:187} plus
minor improvements.  The mapping from effective temperature to SpT
follows the procedure outlined in \citet{Beuermann_etal:612} and
\citet{Baraffe+Kolb:613}. This is not entirely self--consistent as our
code still uses grey atmospheres.

At first we performed a number of calculations for Pop.~I stars with
1.2, 1.6 and $2.0 \, \msun$ and initial ages of 0, 0.25, 0.5, 0.75
and 1 of their MS-lifetime, 
using a switching mass of $M_{\rm sw} = 1 \, \msun$.
As shown in Fig.~\ref{fig:zz}, more massive stars of the same
evolutionary phase and further evolved stars of the same mass
will both lead to larger equilibrium radii after the \TTMT.

Single star radii can easily be converted into orbital periods by
equating to the corresponding Roche radius (with a primary mass given
by $M_1$).
This allows us to compare our tracks with the observed parameters 
for AE~Aqr, $P_{\rm orb} = 9.88 \, {\rm hr}$, SpT = K5 and 
$M_2 = 0.57 \, \msun$. 
For the remaining calculations in this section (shown in Figs.\
\ref{fig:zz_tem} and \ref{fig:zz_tep}) a switching mass of 
$0.7 \, \msun$ was used. 
We varied this value slightly for the purpose of testing, but it
turned out to have little influence on the fitting of final system
parameters.

We obtained four approximate solutions for the initial parameters of
AE~Aqr $(M_{2i}/M_\odot$, $X_{ci}$, $M_2/M_\odot)$:
S1 (1.8, 0.18, 0.6), S2 (1.9, 0.21, 0.6), S3 (2.1, 0.23, 0.54) and
S4 (2.3, 0.247, 0.5),
which provided the starting point for the full binary calculations
in the next section.
A decreasing amount of initial core hydrogen fraction $X_{ci}$
indicates a donor that is advanced futher in its MS evolution. 
The tracks for S3 and S4 are shown in Figs.~\ref{fig:zz_tem} and 
\ref{fig:zz_tep} as the evolution of spectral type over secondary mass
and orbital period respectively.

The (admittedly crude) modelling presented in this section leads to an
overall fitting end state of the system, i.e.\ a reasonable model for
AE~Aqr today. However the preceding transition phase is poorly
modelled. As a general trend derived from comparing S1-S4, lower
initial secondary masses require the donor to be a bit more evolved,
and lead to a slightly higher current donor mass in AE~Aqr.  We can
therefore expect initial values around $M_2 \sim 2
\, \msun$ to provide progenitors for AE~Aqr.

\section{Full binary calculations}

\begin{table}
 \caption[ ]
 {Summary of system parameters for selected full binary calculation.}
 \label{tab:full}
 \begin{center}
 \begin{tabular}{c c c c c c}
	Name &	$M_{1i} / \msun$ & $M_{2i} / \msun$ & $X_{ci}$ & $P_i / {\rm hr}$ & $\eta$
\\ \hline
	B1 &	0.95 & 	2.35 &	0.287 &	17.9 &	0
\\
	B2 &	0.95 &	2.1  &	0.223 &	18.5 &	0
\\
	B3 &	0.95 &	1.8  &	0.190 &	18.1 &	0
\\
	B4 &	0.8  &	2.35 &	0.287 &	17.5 &	0.1
\\
	B5 &	0.6  &	1.6  &	0.560 &	10.8 &	0.3
\\	
	
	B6    &	0.6  &	1.6  &	0.087 &	18.6 &	0.3
\\
 \end{tabular}
 \end{center}
\end{table}

There are two major effects suppressed by the single star treatment in the
previous section: 
\begin{enumerate}
    \item For identical initial donor stars, different initial WD
          masses and thus initial mass ratios can produce very
	  different mass transfer rate histories, which are
	  represented too simplistically by
	  eq.(\ref{switch}).
    \item The growth of the WD mass is ignored in converting
	  radii to orbital periods.
\end{enumerate}    
Therefore we have computed a range of tracks in a more
self--consistent way. The mass transfer rate is calculated
according to the current orbital parameters (`full binary
evolution'). The code used for this work is still based on
the \citet{Kolb+Ritter:187} version of Mazzitelli's stellar evolution.
The main challenge in computing the evolutions through and beyond a
phase of \TTMT\ lies in a proper treatment of numerical and physical 
instabilities.

\begin{figure*}
 \begin{center}
  \centerline{\includegraphics[clip,angle=90,width=0.95\linewidth]{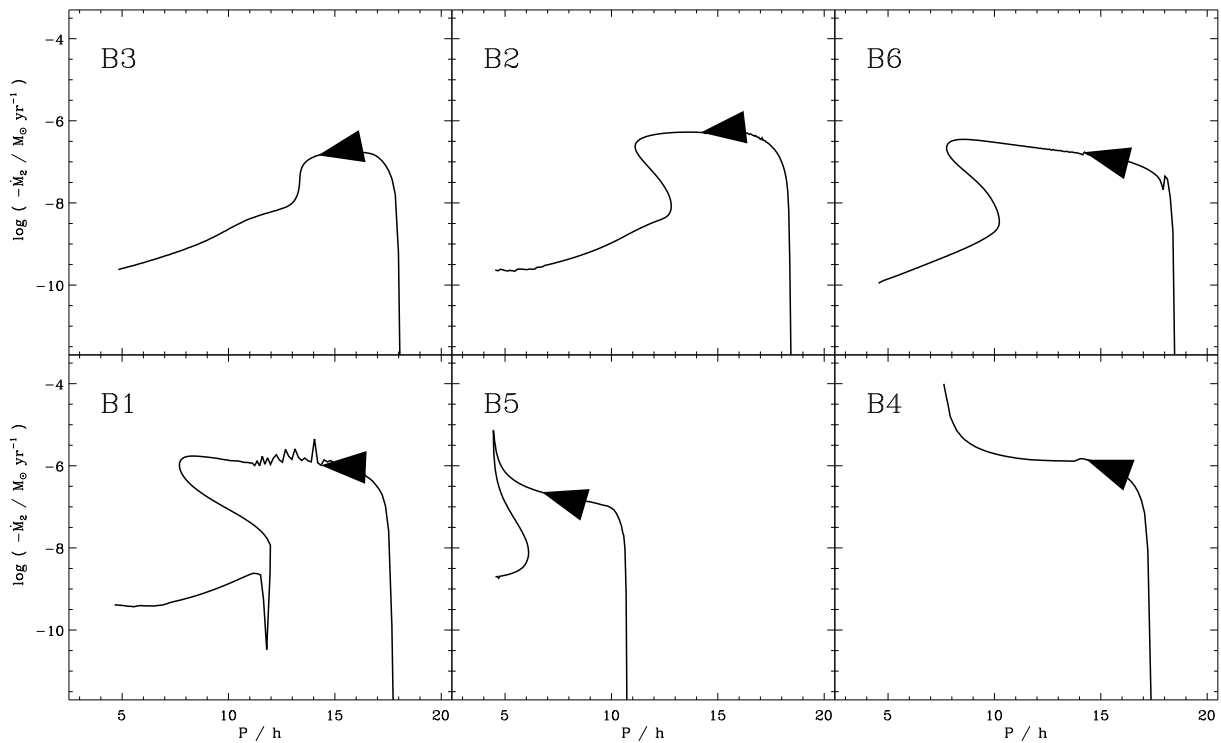}}
  \caption{Possible transitions at the end of \TTMT\ 
shown as the evolution of mass transfer rate over orbital period. 
Each panel is labelled according to system parameters given in Table 
\ref{tab:full}.
The features encountered include a simple drop in mass transfer rate
(B3), a nose-like overhang (e.g.\ B6), up to cases suffering from an
incipient delayed dynamical instability (DDI), countered just in time
by the reversal of the mass ratio in one case (B5).  }
  \label{fig:types} 
 \end{center}
\end{figure*}

We present here the results for six tracks, whose parameters are given
in Table \ref{tab:full}. In all cases the mass accretion efficiency
$\eta$ defined in eq.(\ref{eta}) is kept constant throughout the
evolution. Conservative mass transfer ($\eta = 1$) is not considered
(see  Sect.~2).  Figure  \ref{fig:types}  illustrates the  variety  of
post--\TTMT\  transitions mentioned  in the  discussion of  the single
star results above. Track B3 shows  just a small bump of enhanced mass
transfer.  Increasingly S-shaped period loops appear  in B2, B6, and
B1. Tracks B5 and B4 show  the beginning of an apparent runaway caused
by  a delayed dynamical  instability (DDI);  here extreme  mass ratios
coincide with sufficiently deep transient outer convection zones. None
of these cases, including the dip in the mass transfer rate at the end
of   the   \TTMT\   found   in   B1,  is   adequately   described   by
eq.(\ref{switch}).

  The dominant parameter governing the behaviour along the sequence
  B3--B2--B6--B1--B5--B4 is the initial mass ratio: larger values
  correspond to stronger thermal instability.
  In addition the choice of $\eta$ (modifying the stability criterion)
  and the evolutionary state of the donor (e.g.\ B5--B6)
  can also have a strong impact on the evolutionary curves.

  In some of the curves, particularly B1, a significant amount of
  numerical noise is apparent during the early high mass transfer rate
  phase. This is connected with the growth of an outer convective
  envelope caused by the rapid mass loss. As donors of these masses
  have initially radiative envelopes, very thin convective layers are
  present at the onset of this transition. This creates substantial
  numerical difficulties as minute discontinuities in the evolution of
  the star's radius will be reflected in erratic jumps of the mass
  transfer rate. As time proceeds, the average thermal reaction of the
  star can be seen to pass right through the average of the scatter in
  $\dot{M}_2$: various tests with enhanced resolution and reduced
  timesteps have confirmed that the subsequent structure of the star 
  is unaffected from these problems, which are localized in layers 
  close to the surface and thus have quite short thermal relaxation
  timescales.

\begin{figure}
 \begin{center}
  \centerline{\includegraphics[clip,width=0.95\linewidth]{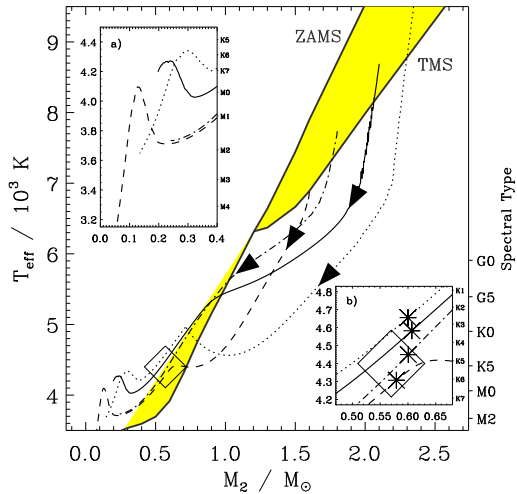}}
  \caption{Tracks of models B1 (dotted), B2 (full), B3 (dash-dotted),
    and B6 (dashed)
    in the plane of effective temperature and 
    secondary mass. The diamond marks the position of AE~Aqr,
    the shaded area framed by thick lines indicates the MS.
    The insets provide a zoomed view of the end stages (a) and 
    AE~Aqr's position today (b).
    In the latter, asterisks are used to mark the position along each
    track where the observed current mass ratio of AE~Aqr is reached,
    showing a fair agreement between the system and the proposed
    progenitor models.
}
  \label{fig:tem}
 \end{center}
\end{figure}
\begin{figure}
 \begin{center}
  \centerline{\includegraphics[clip,width=0.95\linewidth]{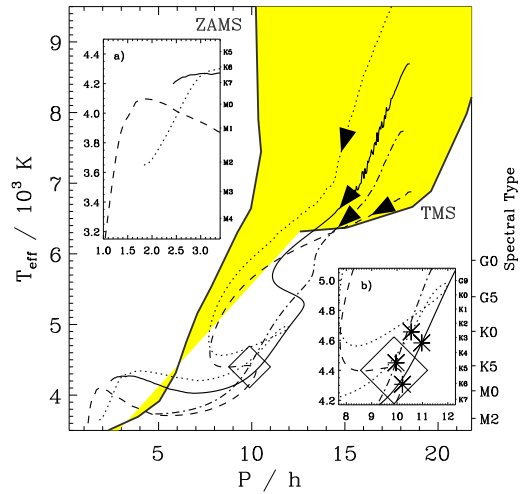}}
  \caption{Tracks of models B1 (dotted), B2 (full), B3 (dash-dotted), 
    and B6 (dashed) in the plane of effective temperature and orbital
    period.
    As in Fig.~\ref{fig:zz_tep} above, the position of AE~Aqr and the
    location of the MS are indicated.
    Similarly insets provide a zoomed view of the end stages (a) 
    and the AE~Aqr position today (b), in which once again
    asterisks (marking the observed current mass ratio of AE~Aqr)
    confirm the successful modelling.
} 
  \label{fig:tep}
 \end{center}
\end{figure}

Here we discuss tracks B1-B3 and B6. Figure \ref{fig:tem} shows
remarkable similarities to Fig.~\ref{fig:zz_tem}, except for the high
initial mass ratio cases (B1 \& B6) which evolve in a much wider arc
back to the mass and SpT of AE~Aqr. 
Inset (a) shows a zoomed part of the end region of the tracks
(discussed in the next section), while inset (b) focusses on the region
near the current position of AE~Aqr (again marked by an open diamond).
Asterisks mark the position along each track where the observed mass
ratio (presumably the best determined observable quantity besides
orbital period) is matched.
Given the uncertainties of spectral type calibration in the models and
observational mass determinations, all four models qualify as 
potential progenitors.
Not suprisingly the evolution vs orbital period shown in
Fig.~\ref{fig:tep} has very different features from the corresponding
Fig.~\ref{fig:zz_tep}: the loops and wiggles in the full binary plot
correspond to the S-shaped curves shown in Fig.~\ref{fig:types}. 
Insets (a) and (b) again zoom in on regions of special interest,
showing the rather good agreement with the observed values of AE~Aqr. 
The jitter visible at the beginning of especially B2 results 
as discussed above
from numerical difficulties at the onset of mass transfer from
an outer convective envelope. Again the strong influence of a large
initial mass ratio on the curves of B1 and B6 can be seen.

All tracks appear very non--MS like after passing through AE~Aqr,
which is a consequence of their advanced nuclear evolution. 
As listed in Table \ref{tab:full} and obvious from Fig.~\ref{fig:tep},
the remaining core hydrogen content at the onset of mass transfer
decreases with initial donor mass (B1--B2--B3--B6) in order to get
near the current position of AE~Aqr.

The full binary calculations in this section confirm the single star
results for a range of suitable progenitors for AE~Aqr. 
Evolutionary tracks for a variety of models do pass through the
currently observed orbital period, spectral type, mass ratio and even
individual masses $M_1$ and $M_2$.
 
So far we have not used the additional constraints
from other observational properties of AE~Aqr. These
distinguish between different scenarios.

\section{Properties of CVs descending from supersoft binaries}

\subsection{The WD spin}

\begin{figure}
 \begin{center}
  \centerline{\includegraphics[clip,width=0.95\linewidth]{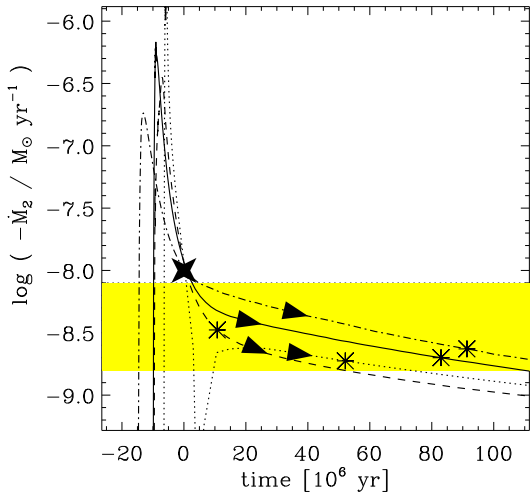}}
  \caption{Time evolution of the mass transfer rate of the
    model sequences B1 (dotted), B2 (full), B3 (dash-dotted),
    and B6 (dashed) shown in the previous figures.
    After a brief ($\la 10^7 {\rm yr}$) phase of high mass transfer
    rates, the system settles towards its current rate. Each track reaches
    the current mass ratio $q = 0.64$ at the position marked by
    an asterisk within the shaded region which covers the
    estimated range of the current mass transfer rate.
    The time axis is offset such that all tracks pass through the
    beginning of the spindown  phase (marked by the spiked cross and
    assumed to start at $10^{-8} \, \msun \, {\rm yr}^{-1}$) at time
    $t = 0$. 
    Larger values of the inital mass ratio $q_i$ accelerate the
    decline and thus shorten the delay between spin--up and the
    currently observed propeller state, leaving model B6 as the most
    suitable progenitor system.}
 \label{fig:spindown}
 \end{center}
\end{figure}

We first consider the current accretion state in AE~Aqr.  The rapidly
rotating WD ($P_{\rm spin} = 33 \, {\rm s}$) is spinning down on a
timescale of $\sim 10^{7} \, {\rm yr}$, consistent with powering the
mass expulsion in the propeller phase \citep{Wynn_etal:624}.  As the
timescale for the spin to reach equilibrium with the mass transfer
rate is $\la 10^7$~yr, the high mass transfer rate which spun up the
WD must have ended only about $10^7 \, {\rm yr}$ in the past. Figure
\ref{fig:spindown} compares the mass transfer rate evolution 
for the four full binary models (B1--B3, B6).
Clearly all cases predict the right $\dot{M}_2$, but all except one (B6)
take too much time to reach that position. The spindown properties
therefore clearly favour a large initial mass ratio for the
progenitor of AE~Aqr, and some fractional mass accumulation (here an
average of $0.3$) on to the WD during the supersoft \TTMT\ phase.


\begin{figure}
 \begin{center}
  \centerline{\includegraphics[clip,width=0.95\linewidth]{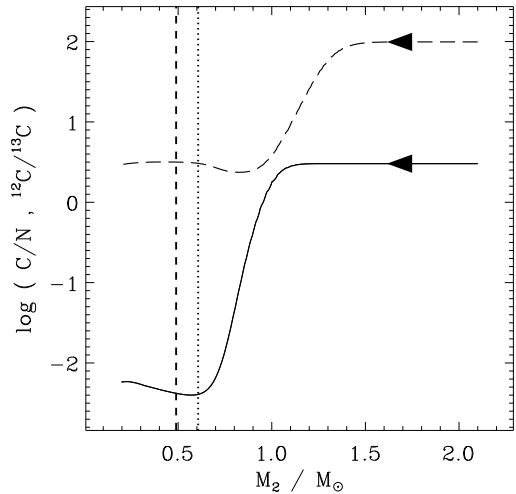}}
  \caption{Change of surface chemical composition in sequence B2:
The system evolves from right to left. After an initial phase of
constant (solar) composition, the ratios of both C/N (lower curve) and
C-12/C-13 (upper curve, dashed) drop severely as the outer
convective envelope reaches down into previously CNO--burning regions. 
Vertical lines mark the position where the sequence reaches the
orbital period (dashed) and mass ratio (dotted) of AE~Aqr
respectively.
} 
 \label{fig:surfratio}
 \end{center}
\end{figure}
\begin{figure}
 \begin{center}
  \centerline{\includegraphics[clip,angle=90,width=0.95\linewidth]{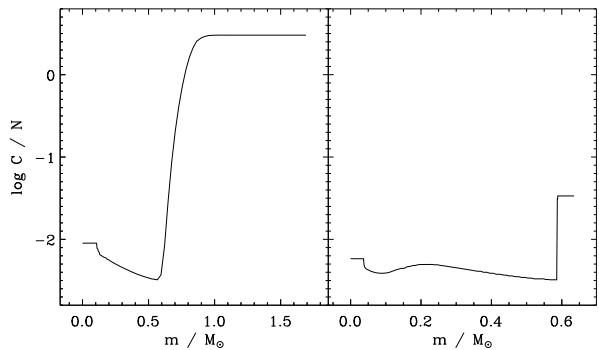}}
  \caption{Internal C/N ratio profile taken from two models of sequence
B3. The lefthand panel shows the situation at $M_2 = 1.69 \, \msun$
during \TTMT, while the model in the righthand panel at
$M_2 = 0.69 \, \msun$ is a close representation of AE~Aqr now. 
Note the reduction of the surface ratio by a factor of $\sim 120$
once mass loss has removed the solar composition layers
completely.}
  \label{fig:zzq}
 \end{center}
\end{figure}

\subsection{Abundances}
Another potential way to distinguish between the models lies
in their different initial masses and central hydrogen fractions.
If we combine the observations of the extreme UV line ratios of
{\sc Civ} and {\sc Nv} with the apparent oversize of the donor
for its spectral type (long orbital period), the natural explanation
is chemical evolution of the
stellar interior. Given the complex
convective history and the different initial core masses and
compositions of our four cases, we can construct e.g.\ an evolutionary
sequence of characteristic surface elements as shown in
Fig.~\ref{fig:surfratio}. Despite the loss of more than a solar mass
from the donor, solar composition is still maintained 
for a long time in sequence B2. 
Later however the ratios of ${}^{12}$C to ${}^{13}$C and C to
N decrease by more than one or two orders of magnitude
respectively. Vertical lines mark the point in the evolution
where the orbital period and mass ratio of AE~Aqr are matched: clearly
this sequence provides more than enough C depletion to explain the UV
observations of \citet{Jameson_etal:1019}. Very similar results are
found in the other cases. Bearing in mind that B2 was not a particularly
strong case of \TTMT, it seems very likely that chemical anomalies
like these are quite common in post--\TTMT\ systems.

In order to understand the origin of this specific form of change in
the C/N ratio, we take a closer look at the internal structure of
sequence B3 in Fig.~\ref{fig:zzq}. Mass transfer in this sequence
started with a $1.8 \, \msun$ donor having only 19\% hydrogen left in
the core. The left panel of Fig.~\ref{fig:zzq} shows the internal
profile of C/N shortly after filling the Roche lobe, with mass $M_2 =
1.69 \, \msun$. The initial solar composition has been modified
severly by nuclear burning and subsequent mixing in the central $\la
1\msun$. Although substantial energy generation by CNO burning is
limited to the very central regions (especially the convective core
plateau in the inner $\sim 0.1 \, \msun$), the relatively long
duration of the previous MS lifetime has allowed the nuclear reactions
to change the composition much further out. The temperature--dependent
CNO equilibrium ratios have been reached out to $0.6\msun$, whereas
the slower nuclear reactions beyond that have only started to reduce
the amount of C. Once \TTMT\ sets in however, this chemical profile
remains frozen (everything happens within $\le 10^{7} \, {\rm yr}$,
cf.\ Fig.~\ref{fig:spindown}). Material is removed layer by layer from
the top, until finally an outer convection zone starts to penetrate
the chemically modified region, or the region itself is laid bare
already. The right panel of 
Fig.~\ref{fig:zzq} 
illustrates this, when only $M_2 = 0.69 \, \msun$ is left. The C/N
ratio in the small surface convection zone is more than a factor of
100 below solar, i.e.\ even the weakest \TTMT\ in case B3 has no
trouble explaining the anomalous composition of AE~Aqr.



\subsection{Evolutionary endpoints: the AM~CVn systems}
The later stages and endpoints of the evolutionary scenarios found for
AE~Aqr are connected to the advanced initial stage of nuclear burning
on the MS.  As the insets (a) in Figs.\ \ref{fig:tem} and
\ref{fig:tep} show, the most extreme case B6 (and quite similar B3,
which has been stopped around $P_{\rm orb} = 5 \, {\rm hr}$) reach an
orbital period of 60 min or below.  In this state the donors consist
largely of helium produced during detached MS burning.
Thus these systems qualify as AM~CVn systems or at least intermediate
cases such as V485~Cen \citep{Augusteijn_etal:687}.  An almost
identical channel of forming ultra-compact binaries has recently been
studied by \citet{Podsiadlowski_etal:866}, focussing on initial donor
masses below $1.4 \, \msun$ and binary population synthesis aspects.

\subsection{Other properties}
\begin{enumerate}
\item The upper limit for the donor mass that potentially leads to
      similar evolutionary tracks is determined by the occurence of a
      DDI, and the requirement of a strong braking mechanism at the
      end of the \TTMT\ supersoft phase. Both are extremely sensitive
      to the amount of mass accreted during the high $\dot{M}_2$
      episode, which certainly needs further investigation. We suggest
      that masses        
      up to $2 \, \msun$ should be considered.
      
\item After passing the state exemplified by AE~Aqr, all donor stars
      in our models have spectral types too late for their orbital
      periods. Hence post--\TTMT\  systems provide a natural
      explanation for the apparently evolved state of many CVs with
      periods above $6 \, {\rm hr}$ \citep{Baraffe+Kolb:613}.
      \citet{Podsiadlowski_etal:866} claim
      agreement between the properties of their population and the
      observed SpT distribution in CVs \citep{Beuermann_etal:612}.
      
\item A simple argument \citep{King+Schenker:891} implies a
      substantial number of evolved systems among CVs: at the
      borderline between systems brought into contact by some form of
      braking and those by nuclear evolution there will always be
      donors which have almost finished their MS phase, but are still
      kept at short orbital periods by a \TTMT\ phase followed by
      normal angular momentum loss driven evolution.  

\item Most importantly, AE~Aqr seems to be living proof that this kind
      of evolution does exist. If (as we strongly suspect; see below) a large number of
      systems with similar properties have passed through such a phase, the
      same group of systems may account for short-period supersoft binaries,
      AM~CVns, V485~Cen--type close binaries, and a substantial fraction of
      CVs. Although these probably include unusual ones like AE~Aqr or
      V1309~Ori, many of them may appear fairly normal.

\item If we assume the track of B6 as shown in Fig.~\ref{fig:spindown}
as typical for all supersoft, AM~CVn, and AE~Aqr--like stages, one
would expect roughly equal durations of the supersoft and spindown
phases. The subsequent lifetime as normal CV above the period gap
is about 100 times longer. More difficult to estimate is the
fraction of post--TTMT systems that end up as AM~CVn stars: the
uncertainties of post-common envelope distributions, DDI, and magnetic
braking issues prevent serious quantitative predictions.  Direct
comparison with observed statistics is further complicated by poorly
understood selection effects in these different classes.  

A related question concerns the expected fraction of post--\TTMT\
systems among CVs. We make a simple estimate using recently improved fits to
the Galactic mass function \citep{Chabrier:1070}, which enter as the
distribution of initial donor star masses between $0.1 \, \msun$ and
$2.1 \, \msun$.
The upper limit is set by the onset of delayed dynamical instability
($q_i = 3$) for a white dwarf mass fixed for simplicity at
$0.7 \, \msun$.
Assuming that these systems evolve as normal CVs for $q_i <1$ and via
\TTMT\ for $q_i > 1$ we find that in a steady state about one--third
of current CVs should be post--\TTMT\ systems. 
As a simple correction for the reduced binary fraction at low masses
we have applied a weighting factors of $0.7$ (for the \TTMT\ mass
range) and $0.3$ (for normal CV donor masses). 
While this is obviously a fairly crude estimate, it does strongly
suggest that a substantial fraction of CVs could show signs of an
earlier \TTMT\ phase. 
We note that more exhaustive studies of CV formation
lead to roughly similar results. For example Table 1 from \citet{Kool:188} 
indicates that a large fraction (if not a majority) of zero--age
binaries will undergo thermally unstable mass transfer rather than  
becoming a CV directly. 
\end{enumerate}

\section{Conclusion}

We have shown that the assumption of \TTMT\ in the
recent past of AE~Aqr provides plausible explanations for all of its
current observational properties. 
The system parameters for the best-fitting progenitor model presented
in this paper (B6) are $M_{2i} = 1.6 \, \msun$ (fairly far evolved on
the MS) and $P_i = 18.6 \, {\rm hr}$ with a primary of 
$M_{1i} = 0.6 \, \msun$ which manages to accrete about 30\% of the
total transferred mass upon reaching the current phase of AE~Aqr.
This indicates a relatively large initial mass ratio, and some
fraction of mass accretion during the \TTMT\ phase.  
Although the precise values may change, the idea is very robust, as
the differential results from Sect.~3 \& 4 clearly indicate that it
will be always possibly to end up with an excellent model for AE~Aqr.

The short duration of the \TTMT\ and the transition stage (where
AE~Aqr currently is) implies a large birthrate, and thus suggests a large 
number of systems passing through similar evolutions. We expect
descendants from systems similar to AE~Aqr, and thus from supersoft
binaries, to form a substantial fraction of the currently known
CVs. The contortions of the $-\dot M_2 - P$ curves of Fig.\
\ref{fig:types} suggest that non--magnetic descendant systems may
change between recurrent novae, novalike and dwarf nova behaviour
during and after the transition from \TTMT\ to normal CV
evolution. Descendants with significant WD magnetic fields will also
appear in various guises during these phases. We suggest that the
long--period AM~Her system V1309~Ori
\citep{Schmidt+Stockman:713,Staude_etal:1020} is a descendant of a
supersoft binary, cf.\ \citet{King_etal:896}. This system is
apparently able to synchronise at its unusually long period of 8~hr
because of the drop in mass transfer rate at the end of the \TTMT\
phase. The pulsing supersoft source XMMU~J004319.4+411758 found by XMM
in M31 may be an example of a progenitor still in the \TTMT\ phase
\citep{King_etal:896}. Note that V1309~Ori, the slightly
nonsynchronous polar BY~Cam \citep{Bonnet-Bidaud+Mouchet:10001} and the
intermediate polar TX Col \citep{Mouchet_etal:10000} all show
high {\sc Nv}/{\sc Civ} ratios similar to AE~Aqr. Abundance anomalies possibly
related to stripping of a partially evolved companion were suggested
by \citet{Mouchet_etal:10000} in the latter two cases.
  
We conclude that in spite of its apparent uniqueness, AE~Aqr is only
the first confirmed member of a much larger population of post--supersoft
binaries, constituting a significant fraction of all CVs. A way of
checking for this population is to determine the C/N ratio by
measuring {\sc Civ} 1550 versus {\sc Nv} 1238. A low value here will be strongly
suggestive of descent from a supersoft binary.


\section*{Acknowledgements}

Theoretical astrophysics research at Leicester is supported by a PPARC
rolling grant. ZZ was supported by Royal Society China Joint Project Q760.



\begin{thebibliography}{}

\bibitem[\protect\astroncite{Augusteijn et~al.}{1996}]{Augusteijn_etal:687}
Augusteijn, T., van~der Hooft, F., de~Jong, J.~A., \& van Paradijs, J.: 1996,
\newblock {A\&A} { 311}, 889

\bibitem[\protect\astroncite{Baraffe \& Kolb}{2000}]{Baraffe+Kolb:613}
Baraffe, I. \& Kolb, U.: 2000,
\newblock {MNRAS} { 318}, 354

\bibitem[\protect\astroncite{Beuermann et~al.}{1998}]{Beuermann_etal:612}
Beuermann, K., Baraffe, I., Kolb, U., \& Weichhold, M.: 1998,
\newblock {A\&A} { 339}, 518

\bibitem[\protect\astroncite{Bonnet--Bidaud \& Mouchet}{1987}]{Bonnet-Bidaud+Mouchet:10001}
Bonnet--Bidaud, \& J.M., Mouchet, M., 1987:
\newblock {A\&A}, { 188}, 89

\bibitem[\protect\astroncite{Casares et~al.}{1996}]{Casares_etal:623}
Casares, J., Mouchet, M., Mart{\'{\i}}nez-Pais, I.~G., \& Harlaftis, E.~T.:
  1996,
\newblock {MNRAS} { 282}, 182

\bibitem[\protect\astroncite{Chabrier}{2001}]{Chabrier:1070}
Chabrier, G.: 2001,
\newblock {ApJ} { 554}, 1274

\bibitem[\protect\astroncite{de~Jager et~al.}{1994}]{Jager_etal:1018}
de~Jager, O.~C., Meintjes, P.~J., O'Donoghue, D., \& Robinson, E.~L.: 1994,
\newblock {MNRAS} { 267}, 577

\bibitem[\protect\astroncite{de Kool}{1992}]{Kool:188}
de Kool, M.: 1992,
\newblock {A\&A} { 261}, 188

\bibitem[\protect\astroncite{Jameson et~al.}{1980}]{Jameson_etal:1019}
Jameson, R.~F., King, A.~R., \& Sherrington, M.~R.: 1980,
\newblock {MNRAS} { 191}, 559

\bibitem[\protect\astroncite{King}{1988}]{King:19}
King, A.~R.: 1988,
\newblock {QJRAS} { 29}, 1

\bibitem[\protect\astroncite{King et~al.}{2002}]{King_etal:896}
King, A.~R., Osborne, J.~P., \& Schenker, K.: 2002,
\newblock {MNRAS} { 329}, L43

\bibitem[\protect\astroncite{King \& Ritter}{1999}]{King+Ritter:610}
King, A.~R. \& Ritter, H.: 1999,
\newblock {MNRAS} { 309}, 253

\bibitem[\protect\astroncite{King \& Schenker}{2002}]{King+Schenker:891}
King, A.~R. \& Schenker, K.: 2002,
\newblock in B.~T. G{\"a}nsicke, K. Beuermann, \& K. Reinsch (eds.),
  {\em The Physics of Cataclysmic Variables and Related Objects},
  ASP Conference Series Vol.~261, San Francisco, 233--241

\bibitem[\protect\astroncite{King et~al.}{2001}]{King_etal:640}
King, A.~R., Schenker, K., Kolb, U., \& Davies, M.~B.: 2001,
\newblock {MNRAS} { 321}, 327

\bibitem[\protect\astroncite{Kolb \& Ritter}{1992}]{Kolb+Ritter:187}
Kolb, U. \& Ritter, H.: 1992,
\newblock {A\&A} { 254}, 213

\bibitem[\protect\astroncite{Mauche et~al.}{1997}]{Mauche_etal:661}
Mauche, C.~W., Lee, Y.~P., \& Kallman, T.~R.: 1997,
\newblock {ApJ} { 477}, 832

\bibitem[\protect\astroncite{Mazzitelli}{1989}]{Mazzitelli:197}
Mazzitelli, I.: 1989,
\newblock {ApJ} { 340}, 249

\bibitem[\protect\astroncite{Mouchet et~al.}{1990}]{Mouchet_etal:10000} 
Mouchet, M., Bonnet--Bidaud, J.M., Hameury, J.M., \& Acker, A.: 1990, 
\newblock in {\em Evolution in Astrophysics: IUE Astronomy in the Era
of New Space Missions}, ESA, 423--426 

\bibitem[\protect\astroncite{Podsiadlowski et~al.}{2001}]{Podsiadlowski_etal:866}
Podsiadlowski, P., Han, Z., \& Rappaport, S.: 2001,
\newblock {MNRAS} submitted (astro--ph/0109171)

\bibitem[\protect\astroncite{Pylyser \& Savonije}{1988a}]{Pylyser+Savonije:587}
Pylyser, E. \& Savonije, G.~J.: 1988a,
\newblock {A\&A} { 191}, 57

\bibitem[\protect\astroncite{Pylyser \& Savonije}{1988b}]{Pylyser+Savonije:588}
Pylyser, E. H.~P. \& Savonije, G.~J.: 1988b,
\newblock {A\&A} { 208}, 52

\bibitem[\protect\astroncite{Schenker}{2001}]{Schenker:724}
Schenker, K.: 2001,
\newblock in P. Podsiadlowski, S. Rappaport, A.~R. King, F. D'Antona, \& L.
  Burderi (eds.), {\em Evolution of Binary and Multiple Star Systems},
  ASP Conference Series Vol.~229, San Francisco, 321--332

\bibitem[\protect\astroncite{Schenker \& King}{2002}]{Schenker+King:892}
Schenker, K. \& King, A.~R.: 2002,
\newblock in B.~T. G{\"a}nsicke, K. Beuermann, \& K. Reinsch (eds.), {\em The
  Physics of Cataclysmic Variables and Related Objects}, 
  ASP Conference Series Vol.~261, San Francisco, 242--251

\bibitem[\protect\astroncite{Schmidt \& Stockman}{2001}]{Schmidt+Stockman:713}
Schmidt, G.~D. \& Stockman, H.~S.: 2001,
\newblock {ApJ} { 548}, 410

\bibitem[\protect\astroncite{Sobermann et~al.}{1997}]{Sobermann_etal:584}
Sobermann, G.~E., Phinney, E.~S., \& van~den Heuvel, E. P.~J.: 1997,
\newblock {A\&A} { 327}, 620

\bibitem[\protect\astroncite{Staude et~al.}{2001}]{Staude_etal:1020}
Staude, A., Schwope, A.~D., \& Schwarz, R.: 2001,
\newblock {A\&A} { 374}, 588

\bibitem[\protect\astroncite{van~den Heuvel et~al.}{1992}]{Heuvel_etal:583}
van~den Heuvel, E. P.~J., Bhattacharya, D., Nomoto, K., \& Rappaport, S.~A.:
  1992,
\newblock {A\&A} { 262}, 97

\bibitem[\protect\astroncite{Webbink}{1985}]{Webbink:702}
Webbink, R.~F.: 1985,
\newblock in J.~E. Pringle \& R.~A. Wade (eds.), {\em Interacting Binary
  Stars}, Chapt. 2.2, 39--70, Cambridge Univ. Press, Cambridge

\bibitem[\protect\astroncite{Welsh}{1999}]{Welsh:668}
Welsh, W.~F.: 1999,
\newblock in {\em Annapolis Workshop on Magnetic CVs}, 
  ASP Conference Series Vol.~157, San Francisco, 357--367

\bibitem[\protect\astroncite{Welsh et~al.}{1993}]{Welsh_etal:637}
Welsh, W.~F., Horne, K., \& Gomer, R.: 1993,
\newblock {ApJ} { 410}, L39

\bibitem[\protect\astroncite{Welsh et~al.}{1995}]{Welsh_etal:625}
Welsh, W.~F., Horne, K., \& Gomer, R.: 1995,
\newblock {MNRAS} { 275}, 649

\bibitem[\protect\astroncite{Wynn et~al.}{1997}]{Wynn_etal:624}
Wynn, G.~A., King, A.~R., \& Horne, K.: 1997,
\newblock {MNRAS} { 286}, 436

\end{thebibliography}



%
%

\label{lastpage}
\end{document}